\documentclass{article}

\usepackage{microtype}
\usepackage{graphicx}
\usepackage{subfigure}
\usepackage{booktabs} 

\usepackage{hyperref}

\usepackage{float}
\usepackage{amsmath}
\usepackage{caption}
\usepackage{siunitx}

\floatstyle{plaintop}
\restylefloat{table}

\DeclareMathOperator*{\argmax}{arg\,max}

\usepackage[accepted]{icml2018}

\begin{document}

\twocolumn[
\icmltitle{Generalization Challenges for Neural Architectures in Audio Source Separation}



\icmlsetsymbol{equal}{*}

\begin{icmlauthorlist}
\icmlauthor{Shariq Mobin}{equal,rw,ucb}
\icmlauthor{Brian Cheung}{equal,rw,ucb}
\icmlauthor{Bruno Olshausen}{rw,ucb}
\end{icmlauthorlist}

\icmlaffiliation{rw}{Redwood Center for Theoretical Neuroscience}
\icmlaffiliation{ucb}{University of California Berkeley}

\icmlcorrespondingauthor{Shariq Mobin}{shariqmobin@berkeley.edu}

\icmlkeywords{Machine Learning, ICML}

\vskip 0.3in
]



\printAffiliationsAndNotice{\icmlEqualContribution} 

\begin{abstract}
    Recent work has shown that recurrent neural networks can be trained to separate individual speakers in a sound mixture with high fidelity. Here we explore convolutional neural network models as an alternative and show that they achieve state-of-the-art results with an order of magnitude fewer parameters. We also characterize and compare the robustness and ability of these different approaches to generalize under three different test conditions:  longer time sequences, the addition of intermittent noise, and different datasets not seen during training. For the last condition, we create a new dataset, \emph{RealTalkLibri}, to test source separation in real-world environments. We show that the acoustics of the environment have significant impact on the structure of the waveform and the overall performance of neural network models, with the convolutional model showing superior ability to generalize to new environments.  The code for our study is available at https://github.com/ShariqM/source\_separation.
\end{abstract}

\section{Introduction}

\begin{figure}[t]
    \centering 
    \begin{subfigure}{}
      \includegraphics[scale=0.28]{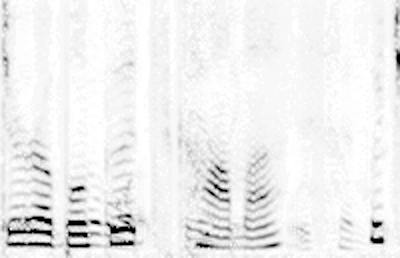}
    \end{subfigure}\hfil 
    \begin{subfigure}{}
      \includegraphics[scale=0.28]{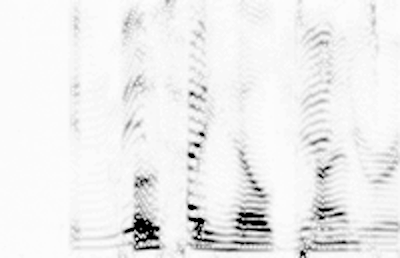}
    \end{subfigure}\hfil 

    \begin{subfigure}{}
      \includegraphics[scale=0.28]{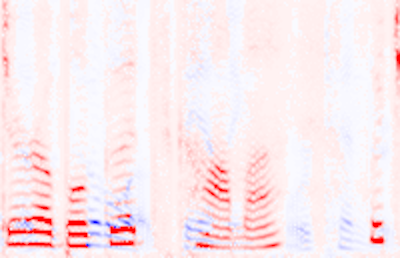}
    \end{subfigure}\hfil 
    \begin{subfigure}{}
      \includegraphics[scale=0.28]{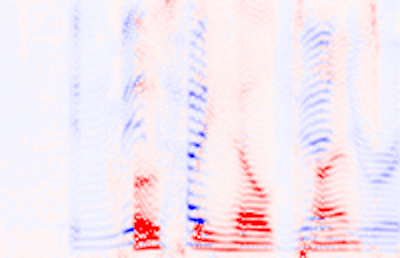}
    \end{subfigure}\hfil 
    \begin{subfigure}{}
      \includegraphics[scale=0.28]{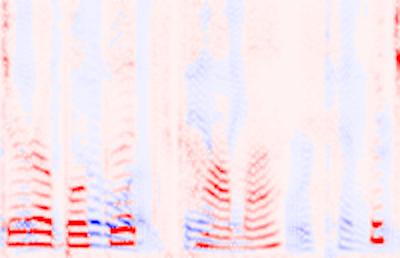}
    \end{subfigure}\hfil 
    \begin{subfigure}{}
      \includegraphics[scale=0.28]{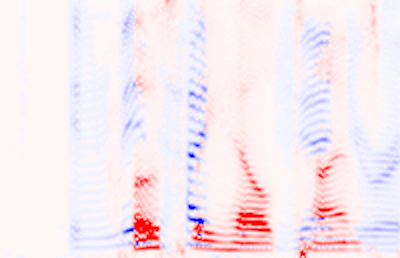}
    \end{subfigure}\hfil 
\setlength{\belowcaptionskip}{-10pt}
\caption{ \emph{Left and right column}: Two examples of source separation using the spectrogram of two overlapped voices as input. \emph{First row}: Spectrogram of the mixture. \emph{Second Row}: The source estimates using the oracle (red and blue). \emph{Third Row}: Source estimates using our method.}
\label{fg:separation_example}
\end{figure}

The sound waveform that arrives at our ears rarely comes from a single isolated source, but rather contains a complex mixture of multiple sound sources transformed in different ways by the acoustics of the environment.  One of the central challenges of auditory scene analysis is to separate the components of this mixture so that individual sources may be recognized.  Doing so generally requires some form of prior knowledge about the statistical structure of sound sources, such as common onset, co-modulation and continuity among harmonic components \cite{bregman1994auditory, darwin1997auditory}. Our goal is to develop a model that can learn to exploit these forms of structure in the signal in order to robustly segment the time-frequency representation of a sound waveform into its constituent sources (see Figure \ref{fg:separation_example}).

The problem of source separation has traditionally been approached within the framework of \textit{computational auditory scene analysis} (CASA) \citep{hu2013unsupervised}. These methods typically rely upon features such as gammatone filters in order to find a representation of the data that will allow for clustering methods to segment the individual speakers of the mixture. In some cases, these features are parameterized to allow for learning \citep{bach2006learning}.  Other approaches use generative models such as factorial Hidden Markov Models (HMMs) to accomplish speech separation or recognition \citep{cooke2010monaural}. Sparse non-negative matrix factorization (SNMF) \cite{le2015sparse} and Bayesian non-parametric models such as \citep{nakano2011bayesian} have also been used. However the computational complexity inherent in many of these approaches makes them difficult to implement in an online setting that is both robust and efficient.

Recently, \citet{hershey2016deep} introduced Deep Clustering (DPCL) which uses a Bi-directional Long short-term memory (BLSTM) \cite{graves2005bidirectional} neural network to learn useful embeddings of time-frequency bins of a mixture. They formulate an objective function which encourages these embeddings to cluster according to their source so that K-means can be applied to partition the source signals in the mixture. This model was further improved in the work of \citet{isik2016single} and \citet{chen2017deep} which proposed simpler end-to-end models and achieved an impressive $\sim$10.5dB Signal-to-Distortion Ratio (SDR) in the source estimation signals.

In this work we develop an alternative model for source separation based on a dilated convolutional neural network architecture \cite{yu2015multi}. We show that it achieves similar state-of-the-art performance as the BLSTM model with an order of magnitude fewer parameters. In addition, our convolutional approach can operate over a streaming signal enabling the possibility of source separation in real-time.

\begin{figure}[ht]
    \centering
    \includegraphics[scale=0.35]{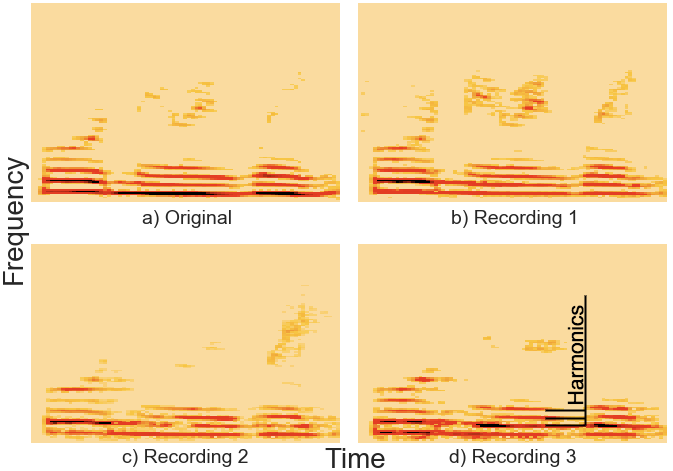}
    \setlength{\belowcaptionskip}{-10pt}
    \caption{\emph{a}: Original recording of a single female speaker from the LibriSpeech dataset; \emph{b,c,d}: Recordings of the original waveform made with three different orientations between computer speaker and recording device.}
    \label{fig:spectrogram_recordings}
\end{figure}

Another goal of this study is to examine how well these different neural network models generalize to inputs that are more realistic. We test the models with inputs containing very long time sequences, intermittent noise, and mixtures collected under different recording conditions, including our \emph{RealTalkLibri} dataset. Success in these more challenging domains is critical for progress to continue in source separation, where the eventual goal is to be able to separate sources regardless of speaker identities, recording devices, and acoustical environments. Figure \ref{fig:spectrogram_recordings} shows three examples of how these factors can affect the spectrogram of the recorded waveform. 
 
While the Automatic Speech Recognition (ASR) community has begun to discuss and address this generalization challenge \cite{vincent2017analysis, hsu2017unsupervised}, there has been less discussion in the context of audio source separation. In vision and machine learning, this issue is usually referred to as \emph{dataset bias} \cite{torralba2011unbiased, tzeng2017adversarial, donahue2014decaf} where models perform well on their training dataset but fail to generalize to new datasets. In recent years the main approach to tackling this issue has been through data augmentation. In the speech community, simulators for different acoustical environments \cite{barker2015third, kinoshita2013reverb} have been leveraged to create more data. 

Here we show that the choice of model architecture alone can improve generalization. Our choice to use a convolutional architecture was inspired by the generalization power of Convolutional Neural Networks (CNNs) \cite{lecun1998gradient, krizhevsky2012imagenet, sigtia2016end} relative to fully connected networks. We compare the performance of our CNN model with the recurrent BLSTM models of previous work and show that while both suffer when tested on reaslistic mixtures under novel recording conditions, the CNN model degrades more gracefully and exhibits superior performance to the BLSTM in this regime.

\begin{figure*}[ht]
    \centering
    \includegraphics[scale=0.38]{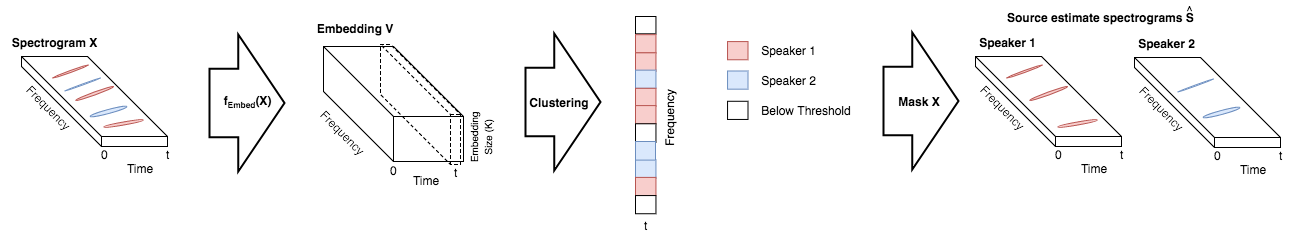}
    \caption{Overview of the source separation process.}
    \label{fg:model_spectrogram_to_embedding}
\end{figure*}

\section{Deep Attractor Framework}

\textbf{Notation:} For a tensor $T \in \mathbf{R}^{A \times B \times C}$: $T_{\cdot, \cdot, c} \in \mathbf{R}^{A \times B}$ is a matrix, and $T_{a,\cdot, c} \in \mathbf{R}^B$ is a vector, and $T_{a,b,c} \in \mathbf{R}$ is a scalar.

\subsection{Embedding the mixed waveform}
\citet{chen2017deep} propose a framework for single-channel speech separation. $x \in \mathbf{R}^{\tau}$ is a raw input signal of length $\tau$ and $X \in \mathbf{R}^{F \times T} $ is its spectrogram computed using the Short-time Fourier transform (STFT). Each time-frequency bin in the spectrogram is embedded into a K-dimensional latent space $V \in \mathbf{R}^{F \times T \times K}$ by a learnable transformation $f(\cdot; \theta)$ with parameters $\theta$:
\begin{eqnarray}
\bar{V} = f(X; \theta) \\
\label{eq:embedding_fn}
V_{f,t,\cdot} = \frac{\bar{V}_{f,t,\cdot}}{|| \bar{V}_{f,t,\cdot} ||_2}
\label{eq:normalize}
\end{eqnarray}

In our work, the embeddings are normalized to the unit sphere in the latent dimension $k$ (eq. \ref{eq:normalize}).

\subsection{Generating embedding labels}

We assume that each time-frequency bin can be assigned to one of the $C$ possible speakers. The Ideal Binary Mask (IBM), $\bar{Y} \in \mathbf{\{0,1\}}^{F \times T \times C}$, is a one-hot representation of this classification for each time-frequency bin:
\begin{eqnarray}
\bar{Y}_{f,t,c} &=& 
    \begin{cases}
        1, & \text{if} \hspace{2mm} c =  \underset{c'}{\argmax} (S_{f,t,c'}) \\
        0, & \text{otherwise}
    \end{cases}
\label{eq:IBM}
\end{eqnarray}

where $S \in \mathbf{R}^{F \times T \times C}$ is the supervised source target spectrogram. We estimate $\bar{Y}$ by a mask $M \in (0,1)^{F \times T \times C}$ computed from the spectrogram $X$.

To prevent time-frequency embeddings with negligible power from interfering, the raw classification tensor $\bar Y$ is first masked with a threshold tensor $H \in \mathbf{R}^{F \times T} $. The threshold tensor removes time-frequency bins which are below a fraction $0 < \alpha < 1$ of the highest power bin present in $X$:
\begin{eqnarray}
H_{f,t} &=& 
\begin{cases}
0, & \text{if} \hspace{2mm} X_{f,t} < \alpha \hspace{1mm} \text{max}(X) \\
1, & \text{otherwise}
\end{cases}\\
Y_{\cdot, \cdot, c} &=& \bar Y_{\cdot, \cdot, c} \odot H
\label{eq:threshold}
\end{eqnarray}

where $\odot$ denotes element-wise product. 

\subsection{Clustering the embedding}

An attractor point, $A_c \in \mathbf{R}^{K}$, can be thought of as a cluster center for a corresponding source $c$. Each attractor $A_c$ is the mean of all the embeddings which belong to speaker $c$:
\begin{eqnarray}
A_{c,k} &=& \frac{\sum_{f,t} V_{f,t,k} Y_{f,t,c}}{\sum_{f,t} Y_{f,t,c}}
\label{eq:centroid}
\end{eqnarray}

During training the attractor points are calculated using the thresholded oracle mask, $Y$. In the absence of the oracle mask at test time, the attractor points are calculated using K-means. Only the embeddings which pass the corresponding time-frequency bin threshold are clustered.

Finally the mask is computed by taking the inner product of all embeddings with all attractors and applying a softmax:
\begin{equation}
M_{f,t,c} = \underset{c}{softmax}\Big(\sum_k A_{c, k} V_{f,t,k}\Big)
\label{eq:mask}
\end{equation}

From this mask, we can compute source estimate spectrograms:
\begin{equation}
\hat S_{\cdot, \cdot, c} = M_{\cdot, \cdot, c} \odot X
\label{eq:reconstruct_source}
\end{equation}

which in turn can be converted back to an audio waveform via the inverse STFT. We do not attempt to compute the phase of the source estimates. Instead, we use the phase of the mixture to compute the inverse STFT with the magnitude source estimate spectrogram $\hat S$.

The loss function $\mathcal{L}$ is the mean-squared-error (MSE) of the source estimate spectrogram and the supervised source target spectrogram, $S \in \mathbf{R}^{F \times T \times C}$:
\begin{eqnarray}
\mathcal{L} &=& \sum_{c} || S_{\cdot, \cdot, c} - \hat S_{\cdot, \cdot, c} ||^2_F  \\
\end{eqnarray}

where $||\cdot||_F$ denotes the Frobenius norm. See Figure \ref{fg:model_spectrogram_to_embedding} for an overview of our source separation process.

\subsection{Network Architecture}
A variety of neural network architectures are potential candidates to parameterize the embedding function in Equation \ref{eq:embedding_fn}. \citet{chen2017deep} use a 4-layer Bi-Directional LSTM architecture \citep{hochreiter1997long, schuster1997bidirectional}. This architecture utilizes weight sharing across time which allows it to process inputs of variable length.

By contrast, convolutional neural networks are capable of sharing weights along both the time and frequency axis. Recently convolutional neural networks have been shown to perform state-of-the-art music transcription by having filters which convolve over both the frequency and time dimensions \cite{sigtia2016end}. One reason this may be advantageous is that the harmonic series exhibits a stationarity property in the frequency dimension. Specifically, for a signal with fundamental frequency $f_0$ the harmonics are equally spaced according to the following set: $ \{i * f_0 : i = 2, 3, ..., n\}$. This structure can be seen in the equal spacing of successive harmonics in Figure \ref{fig:spectrogram_recordings}d. 

Another motivation we have for using convolutional neural networks is that they do not incorporate feedback which may allow them to be more stable under novel conditions not seen during training. In the absence of a recurrent memory, filter dilation \citep{yu2015multi} enables the receptive field to integrate information over long time sequences without the loss of resolution. Furthermore, incorporating a fixed amount of future knowledge in the network is straightforward by having a fixed-lag delay in the convolution as we show in Figure \ref{fig:fixed_lag_dilation}. This is similar to fixed-lag smoothing in Kalman filters \cite{moore1973discrete}.

\section{Our Model}

\begin{table*}[h]
\begin{center} 
\begin{tabular}{cccccccccccccc} \toprule
    Layer & 1 & 2 & 3 & 4 & 5 & 6 & 7 & 8 & 9 & 10 & 11 & 12 & 13 \\ \midrule
    Convolution & 3x3 & 3x3 & 3x3 & 3x3 & 3x3 & 3x3 & 3x3 & 3x3 & 3x3 & 3x3 & 3x3 & 3x3 & 3x3 \\ \midrule
    Dilation & 1x1 & 2x2 & 4x4 & 8x8 & 16x16 & 32x32 & 1x1 & 2x2 & 4x4 & 8x8 & 16x16 & 32x32 & 1x1  \\ \midrule
    Residual & False & True & False & True & False & True & False & True & False & True & False & True & False \\ \midrule
    Channels & 128 & 128 & 128 & 128 & 128 & 128 & 128 & 128 & 128 & 128 & 128 & 128 & K \\ \bottomrule
\end{tabular}
\caption{Dilated Convolution Model Architecture} \label{ConvModel}
\end{center}
\end{table*}

\subsection{Dilated Convolution}
\citet{yu2015multi} proposed a dilation function $D(\cdot, \cdot, \cdot; \cdot)$ to replace the pooling operations in vision tasks. For notational simplicity, we describe dilation in one dimension. This method convolves an input signal $X \in \mathbf{R}^G$ with a filter $K \in \mathbf{R}^H$ with a dilation factor $d$:
\begin{equation*}
F_t = D(K, X, t; d) = \underset{dh + g = t}{\sum} K_h X_g
\end{equation*}

\begin{figure}
    \centering
    \includegraphics[scale=.5]{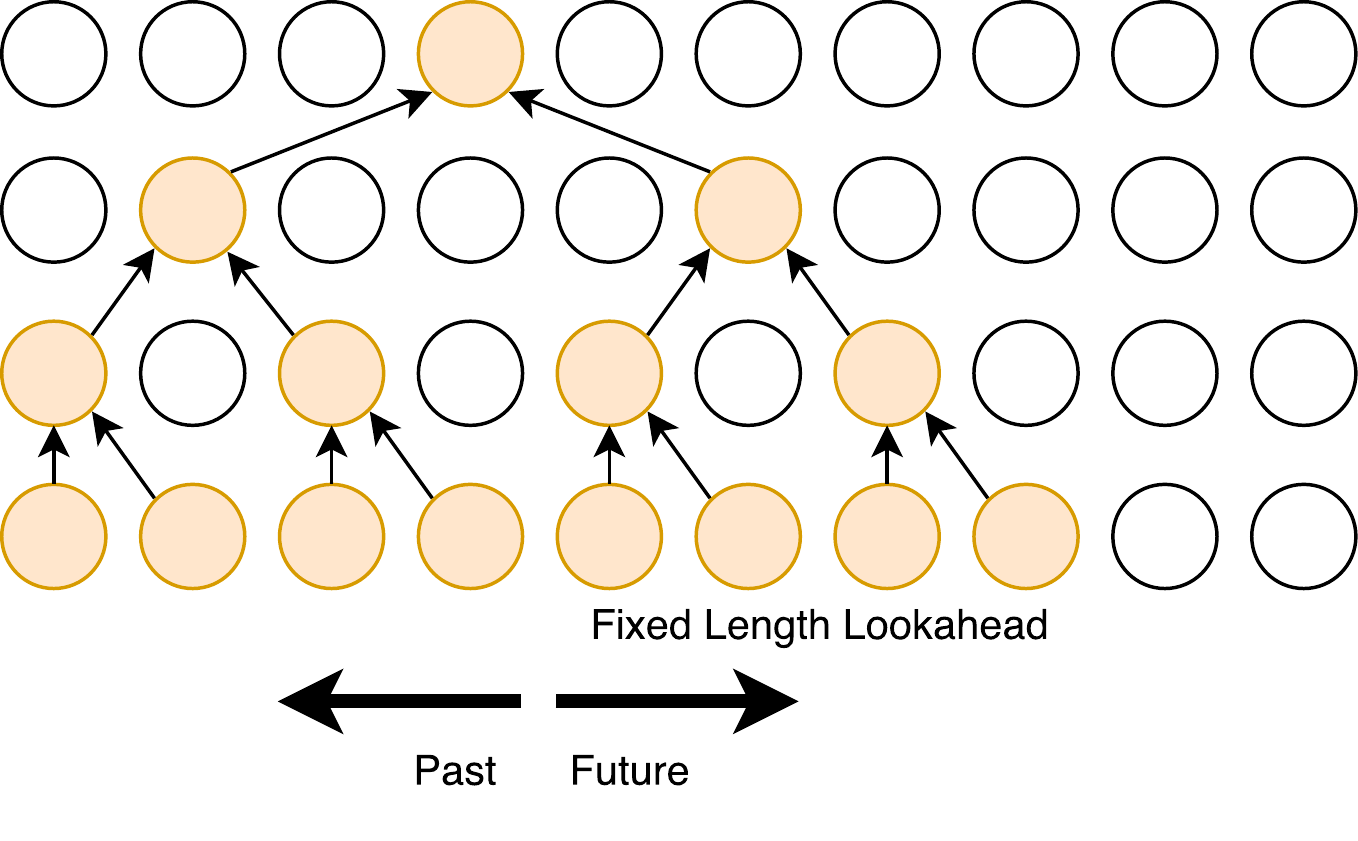}
    \setlength{\belowcaptionskip}{-10pt}
    \caption{One-dimensional, fixed-lag dilated convolutions used by our model with dilation factors of 1, 2 and 4. The bottom row represents the input and each successive row is another layer of convolution. This network has a fixed-lag of 4 timepoints before it can output a decision for the current input.}
    \label{fig:fixed_lag_dilation}
\end{figure}

The input receptive field of a unit $F_t$ in an upper layer of a dilated convolutional network grows exponentially as a function of the layer number as shown in Figure \ref{fig:fixed_lag_dilation}. When applied to time sequences, this has the useful property of encoding long range time dependencies in a hierarchical manner without the loss of resolution which occurs when using pooling. Unlike recurrent networks which must store time dependencies of all scales in a single memory vector, dilated convolutions stores these dependencies in a distributed manner according to the unit and layer in the hierarchy. Lower layers encode local dependencies while higher layers encode longer range global dependencies. Such models have been successfully used for generating audio directly from the raw waveform \citep{oord2016wavenet}.

\section{Datasets}
We construct our mixture data sets according to the procedure introduced in \cite{hershey2016deep}, which is generated by summing two randomly selected waveforms from different speakers at signal-to-noise ratios (SNR) uniformly distributed from -5dB to 5dB and downsampled to 8kHz to reduce computational cost. 

A training set is constructed using speakers from the Wall Street Journal (WSJ0) training dataset \cite{garofalo2007csr} si\_tr\_s.

We construct three test sets:

1) In WSJ0, a test set is constructed identical to the test set introduced in \cite{hershey2016deep} using 18 unheard speakers from  si\_dt\_05 and si\_et\_05.

2) In LibriSpeech, a test set is constructed using 40 unheard speakers from test-clean.

3) In \emph{RealTalkLibri}, we generate more realistic mixture using the procedure described below.

\subsection{\emph{RealTalkLibri} Dataset}
The main motivation for creating this dataset is to record mixtures of speech where the acoustics of the room deform a high quality recording into a more realistic one. While datasets of real mixtures exist, there exists no dataset where the ground truth source waveforms are available, only the transcription of the speakers words are given as target outputs \cite{kinoshita2013reverb, barker2015third}. In order to understand how well our model generalizes to real world mixtures we created a small test dataset for which there is ground truth of the source waveforms. 

\begin{figure}[h]
    \centering
    \includegraphics[scale=0.5]{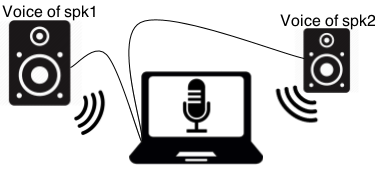}
    \caption{Recording setup diagram.} 
    \label{recording}
\end{figure}

The \emph{RealTalkLibri} (RTL) test dataset is created starting from the test-clean directory of the open LibriSpeech dataset \cite{panayotov2015librispeech} which contains 40 speakers. We first downsampled all waveforms to 8kHz as before. Each mixture in the dataset is created by sampling two random speakers from the test-clean partition of LibriSpeech, picking a random waveform and start time for each, and playing the waveforms through two Logitech computer speakers for 12 seconds. The waveforms of the two speakers are played in separate channels linked to a left and right computer speaker, separated from the microphone of the computer by different distances. The recordings are made with a sample rate of 8kHz using a 2013 MacBook Pro (Figure \ref{recording}). In order to obtain ground truth of the individual speaker waveforms each of the waveforms is played twice, once in isolation and once simultaneously with the other speaker. The first recording represents the ground truth and the second one is for the mixture. To verify the quality of the ground truth recordings, we constructed an ideal binary mask $\bar{Y}$ which performs about as well on the previous simulated datasets, see the Oracle performance in Figure \ref{fg:rtl_results}. The \emph{RealTalkLibri} data set is made up of two recording sessions which each yielded 4.5 hours of data, giving us a total of 9 hours of test data.  The data is available at https://www.dropbox.com/s/4pscejhkqdr8xrk/rtl.tar.gz?dl=0

\section{Experiments}
\subsection{Experimental Setup}
We evaluate the models on a single-channel simultaneous speech separation task. The mixture waveforms are transformed into a time-frequency representation using the Short-time Fourier Transform (STFT) and the log-magnitude features, $X$, are served as input to the model. The STFT is computed with 32ms window length, 8ms hop size, and the Hann window. We use SciPy \cite{jones2014scipy} to compute the STFT and TensorFlow to build our neural networks \cite{abadi2016tensorflow}.

We report our results using a distance measure between the source estimate and the true source in the waveform space. Our distance measure is the signal-to-distortion ratio (SDR) which was introduced in \cite{vincent2006performance} as a blind audio source separation (BSS) metric which is less sensitive to the gain of the source estimate. We compute our results using version 3 of the Matlab bsseval toolbox \cite{fevotte2005bss_eval}. A python implementation of this code is also available online \cite{raffel2014mir_eval} \footnote{\url{https://github.com/craffel/mir_eval/blob/master/mir_eval/separation.py}}.
    
Our network consists of 13 dilated convolutional layers \citep{yu2015multi} made up of two stacks, each stack having its dilation factor double each layer. Batch Normalization \cite{ioffe2015batch} is applied to each layer and residual connections \cite{he2016deep} are used at every other layer, see Table \ref{ConvModel} for details. Our model has a fixed-lag response of 127 timepoints ($\sim$ 1s, see Figure \ref{fig:fixed_lag_dilation}). The output of the network is of dimensionality ($T \times F \times K$), $T$ being the number of output time points, F the number of frequency bins, and $K$ being both the final number of channels and embedding dimensionality. During training $T$ is set to 400 ($\sim$3s), $F$ to 129 (specified by the STFT), $K$ to 20, and $\alpha$ (threshold factor) to 0.6. For evaluation we also use the max() function rather than the softmax()  for computing the mask in equation (\ref{eq:mask}). The Adam Optimizer \cite{kingma2014adam} is used with a piecewise learning rate schedule, boundaries$= [10k, 50k, 100k]$, values $= [1.0, 0.5, 0.1, 0.01]$, and initial learning rate $1\mathrm{e}{-3}$.

We reimplement the DANet of \cite{chen2017deep} with a BLSTM architecture containing 4 layers and 500 hidden units in both the forward and backward LSTM, for a total of 1000 hidden units. We replicated their training schedule using the RMSProp algorithm \cite{tieleman2012lecture}, a starting learning rate of $1\mathrm{e}{-3}$, and exponential decay with parameters: decay\_steps$ = 2000$, decay\_rate$=0.95$.

We calculate an Oracle score using the Ideal Binary Mask (IBM), $\bar{Y}$, using the ground truth source spectrograms (Eq. \ref{eq:IBM}).

\subsection{WSJ0 Evaluation}
We begin by evaluating the models on the WSJ0 test dataset as in \cite{chen2017deep}. Our state-of-the-art results are shown in Table \ref{tb:sota_results}. Our model achieves the best score using a factor of ten fewer parameters than DANet. The DPCL score is taken from \cite{isik2016single} which has a very similar architecture to DANet and therefore a similar number of parameters. Their model has one important difference however, a second neural network is used to enhance the source estimate spectrogram to achieve their result. Our model is still able to exceed its performance without this extra enhancement network. In addition, our model has a fixed window into the future whereas the BLSTM models have access to the entire future. This indicates that a convolution based architecture is better at solving this source separation task with less information in comparison to a recurrent based architecture.

\begin {table}[h]
    \begin{center}
        \begin{tabular}{SSSSS} \toprule
            {Model} & {WSJ0} & {Number of}  \\
            {} & {SDR (dB)} & {Parameters} \\ \midrule
            {DANet (BLSTM)} & {10.48} &  {\num{17114580}} \\ 
            {DPCL* (BLSTM)} & {10.8} &  {?} \\
            {Ours (CNN)} & {\textbf{10.97}} & {\num{1650836}} \\ 
            {Oracle} & {13.49} &  {-}  \\ \bottomrule
        \end{tabular}
    \setlength{\belowcaptionskip}{-10pt}
    \caption{Signal-to-Distortion Ratio (SDR) for two competitor models, our proposed convolutional model, and the Oracle. Our model achieves the best results using significantly fewer parameters. Our score is averaged over 3200 examples. *: SDR score is from \cite{isik2016single}. This model is at an advantage because it has a second enhancement neural network that improved the source estimate spectrogram after masking in addition to the normal BLSTM.} \label{tb:sota_results}
    \end{center}
\end{table}

\subsection{Embeddings}
At test time we do not have access to the labels $\bar{Y}$ so the attractors cannot be computed using equation \ref{eq:centroid}. As in previous work, K-means is employed instead. In order for the attractors to from a good proxy for the K-means algorithm it is important the attractors form dense clusters of embeddings. Without applying any regularization on the attractors we found that this was not the case. The main issue we observed was that embeddings for both speakers in the mixture were largely overlapping with a few embeddings driven extremely far apart in order to drive the attractors apart. This worked well for training but poorly at test time, the solutions found by K-means didn't match the attractors found in training. In order to combat this degeneracy we l2 normalize the embeddings V which is novel and very effective (Eq. \ref{eq:normalize}).

\begin{figure}
    \centering
    \includegraphics[scale=0.23]{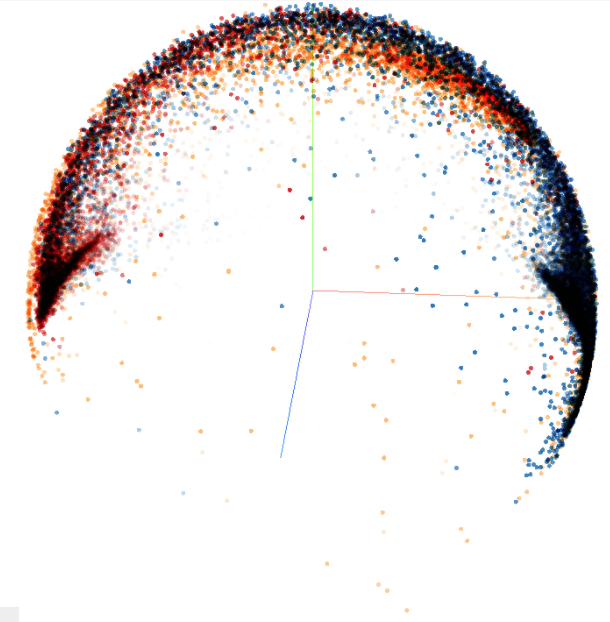}
    \caption{Embeddings for two speakers (red \& blue) over 200 time points, projected onto a 3-dimensional subspace using PCA. The orange points correspond to time-frequency bins where the energy was below threshold (see eq. \ref{eq:threshold}). There are $T \times F = 200 \times 129 = 25800$ embedding points in total. } 
    \label{clusters}
\end{figure}

In Figure \ref{clusters} we visualize the embedding outputs of our model using PCA for a single mixture across $T=200$ timepoints. Each embedding point corresponds to a single time-frequency bin in the mixed input spectrogram. The embeddings are colored in this diagram according to the oracle labelling, red for speaker 1 and blue for speaker 2. Notice that the network has learned to cluster the embeddings according to the speaker they belong too, i.e. there is a high density of red embeddings on the left and similarly for blue embeddings on the right. This structure allows K-means to easily find cluster centers that match the attractors used at training time. 

\begin{figure*}[htb!]
    \centering
    \includegraphics[scale=0.58]{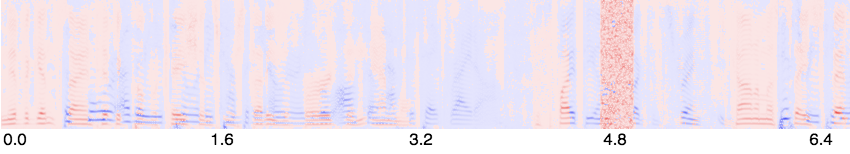}
    \caption{Source separation spectrogram for the noise generalization experiment.} 
    \label{fg:attractor_noise}
\end{figure*}

\subsection{Generalization Experiments}
\subsubsection{Length Generalization}
\begin{figure}[t]
    \centering
    \includegraphics[scale=0.3]{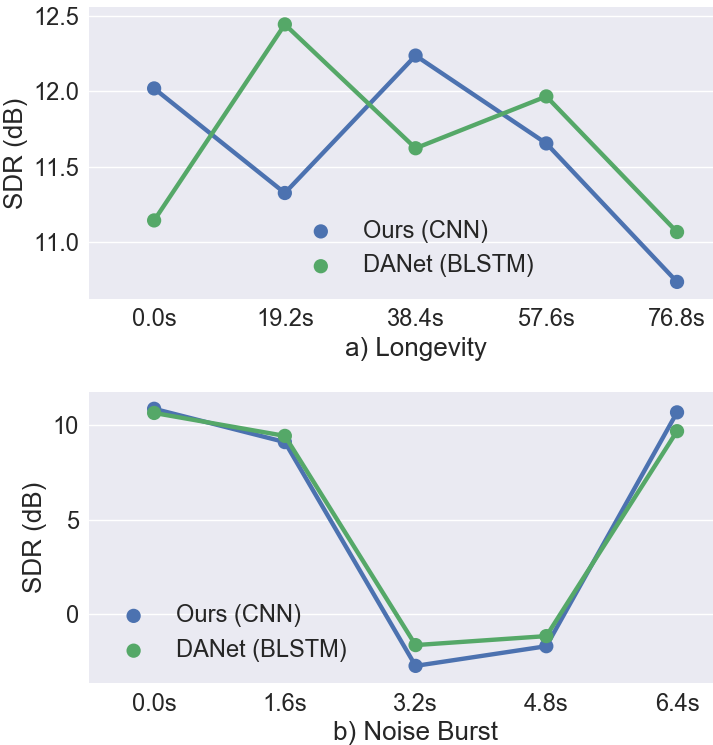}
    \caption{Results of the length and noise generalization experiments. For both plots, we plot the SDR starting at the time specified by the x-axis up until 400 time points ($\sim 3$s) later. Both the BLSTM and CNN are (a) able to operate over very long time sequences and (b) recover from intermittent noise. See Figure \ref{fg:attractor_noise} for the spectrogram in b).} 
    \label{fg:longevity_and_noise}
\end{figure}

In the first experiment we study how well these models work under time-sequences 25x longer than they are trained on, i.e. $T=10000$ ($\sim80$s). Previous work \citep{kaiser2015neural} has indicated that because recurrent architectures incorporate feedback they can function unpredictably for sequence lengths beyond those seen during training. On the other hand, convolutional network architectures do not incorporate any feedback. This is advantageous for processing time sequences of indefinite length because errors cannot accumulate over time. Since a convolutional network is a stationary process in the convolved dimension, we hypothesize this architecture will operate more robustly over sequence lengths much  longer than those seen during training. Our results are shown in Figure \ref{fg:longevity_and_noise}a. Surprisingly, the results indicate that the BLSTM is also able to generalize to sequences of significantly larger length, contrary to our expectations. We discuss possible explanations of this result in the next section. Our CNN model is able to maintain its performance across the long sequence as expected. 

\subsubsection{Noise Generalization}
In the second experiment we are interested in how the models respond to small bursts of input data far outside of the training distribution. We believe the BLSTM model might become unstable as a result of such inputs because its recurrent structure makes it possible for the noise to affect its hidden state indefinitely. We took sequences of length $T=1200$ ($\sim9$s) and inserted white noise for 0.25s in the middle of the mixture to disrupt the models process. Our results are shown in Figure \ref{fg:longevity_and_noise}b. Again, contrary to our belief the BLSTM is very resilient to this noise, the model quickly recovers after the noise passes (last data point). One possible explanation is that the BLSTM is only integrating information over short time scales and therefore ``forgets'' about previous input data after a short number of time steps. We believe this is because when we construct the input for our models we randomly sample a starting time for each waveform. This may force the BLSTM to learn a stationary function since it must be able to separate the mixture with or without information from the past in its hidden state.

\subsubsection{Data Generalization}
\begin{figure}[ht]
    \centering
    \includegraphics[scale=0.28]{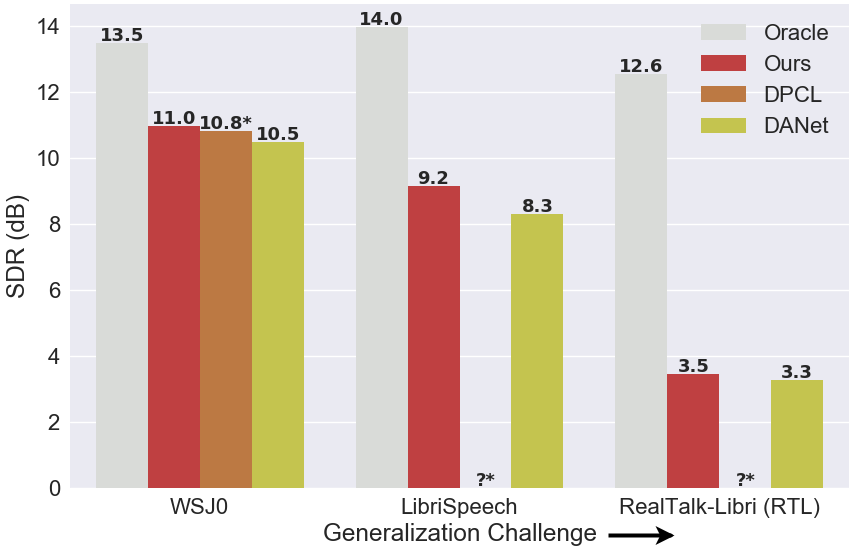}
    \caption{Results of models tested on WSJ0 simulated mixtures, LibriSpeech simulated mixtures, and our RealTalkLibri (RTL) dataset. Our model performs the best on WSJ0, generalizes better to LibriSpeech, but fails alongside the BLSTM at generalizing to the real mixtures of RTL. All models are trained on WSJ0. *: Model has a second neural network to enhance the source estimate spectrogram and is therefore at an advantage. The model wasn't available online for testing against LibriSpeech or RTL.} 
    \label{fg:rtl_results}
\end{figure}

In the final experiment we are interested in how well the models generalize to data progressively farther from their training distribution. We trained all the models on the WSJ0 training set and then tested on the WSJ0 test set, the LibriSpeech test set, and \emph{RealTalkLibri} test set. Our results are shown in Figure \ref{fg:rtl_results}. Our model generalizes quite well from the WSJ0 dataset to the LibriSpeech dataset, only losing $1.8$dB of performance. Unfortunately it degrades substantially, by $7.5$dB, when using the RTL dataset. However, our model still outperforms the DANet model on all datasets. Note that the Oracle performance also degrades by $\sim 1$dB on RTL.

\begin{figure}[ht]
    \centering 
    \begin{subfigure}{}
      \includegraphics[scale=0.55]{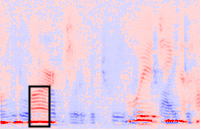}
    \end{subfigure}\hfil 
    \begin{subfigure}{}
      \includegraphics[scale=0.55]{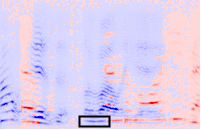}
    \end{subfigure}\hfil 

    \begin{subfigure}{}
      \includegraphics[scale=0.55]{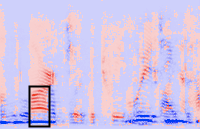}
    \end{subfigure}\hfil 
    \begin{subfigure}{}
      \includegraphics[scale=0.55]{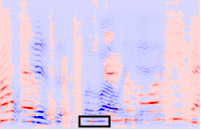}
    \end{subfigure}\hfil 
\caption{\emph{Left and right column}: Two examples of source separation on the \emph{RealTalkLibri} dataset. \emph{Row 1}: Source estimates using the oracle. \emph{Row 2}: The source estimates using our method. The model has difficulty maintaining continuity of speaker identity across frequencies of a harmonic stack (left column) and across time (right column).}
\label{fg:rtl_separation_example}
\end{figure}

In Figure \ref{fg:rtl_separation_example} we visualize the mistakes our network makes under the \emph{RealTalkLibri} dataset. The first example indicates that the model does not have a strong enough bias to the harmonic structure contained in speech, it classifies the frequencies of the fundamental to a different speaker than the harmonic frequencies of that fundamental. The second example indicates that the model also has issues with temporal continuity, the speaker identity of particular frequency bins varies sporadically across time. This indicates that there is still room to improve generalization in these models by modifying model architecture and adding regularization.

\section{Discussion}
Recurrent neural networks have been shown to perform audio source separation with high fidelity. Here we explored using convolutional neural networks as an alternative model. Our state-of-the-art results on the WSJ0 dataset using a factor of ten fewer parameters show that convolutional models are both more accurate and more efficient for audio source separation. Our model has the additional advantage of working online with a fixed-lag response of $\sim 1$sec.

In order to study the robustness of all models we studied their performance under three different conditions: longer time sequences, intermittent noise, and datasets not seen during training. Our results in the length and noise generalization experiments indicate that the BLSTM learns to behave much like a stationary process in the temporal dimension. We do not observe any substantial degradation in performance after it has been perturbed with noise. It also performs consistently on sequences which are significantly longer than those seen during training. 

 On the other hand, we get this stationarity property for free with our convolutional model. This further motivates our network architecture in Figure \ref{fig:fixed_lag_dilation} which, by design, integrates only local information from the past and future.

In the final experiment we showed that our convolutional neural network also generalized better to both the LibriSpeech dataset and the \emph{RealTalkLibri} dataset we introduced here. Models which are robust to new datasets as well as the deformations caused by the acoustics of different environments are critical to progress in audio source separation. Our \emph{RealTalkLibri} dataset complements other real-world speech datasets \cite{barker2015third, kinoshita2013reverb} by additionally providing approximate ground truth waveforms for the mixture which is currently not available.

Looking forward, we aim to improve the generalization ability on examples such as those shown in Figure \ref{fg:rtl_separation_example} by introducing a training set for \emph{RealTalkLibri}, developing more robust model architectures, introducing regularizers for the structure of speech, and creating powerful data augmentation tools. We also believe models which can operate under an unknown number of sources is of utmost importance to the field of audio source separation.

\newpage
\bibliography{example_paper}
\bibliographystyle{icml2018}

\end{document}